\documentclass[journal,12pt,onecolumn]{IEEEtran}
\usepackage{cite}
\usepackage{amsmath,amssymb,amsfonts}
\usepackage{algorithmic}
\usepackage{graphicx}
\usepackage{textcomp}
\usepackage{comment}
\usepackage{atbegshi}
\AtBeginDocument{\AtBeginShipoutNext{\AtBeginShipoutDiscard}}

\def\BibTeX{{\rm B\kern-.05em{\sc i\kern-.025em b}\kern-.08em
    T\kern-.1667em\lower.7ex\hbox{E}\kern-.125emX}}

\newtheorem{proposition}{Proposition}[section]

\newcommand{\PP}{\mathbb{P}} 

\newcommand{\ee}{{\mathrm e}}

\newcommand{\Cc}{{\cal C}}

\def\ben{\begin{enumerate}}
\def\beq{\begin{equation}}
\def\beqa{\begin{eqnarray}}
\def\bit{\begin{itemize}}
\def\een{\end{enumerate}}
\def\eeq{\end{equation}}
\def\eeqa{\end{eqnarray}}
\def\eit{\end{itemize}}

\def\non{\nonumber\\}

\newcommand{\pf}{\noindent{\bf Proof:~}}
\newcommand{\qedsymb}{\hfill{\rule{2mm}{2mm}}}

\newif\ifforarxiv
\forarxivtrue

\ifforarxiv

\begin{document}

\title{Generalized Quantum-assisted Digital Signature}
\author{Alberto Tarable, \IEEEmembership{Member, IEEE},
Rudi Paolo Paganelli, Elisabetta Storelli, Alberto Gatto, \IEEEmembership{Member, IEEE} and Marco Ferrari
\IEEEmembership{Member, IEEE}.}
\thanks{A. Tarable, R. P. Paganelli and M. Ferrari are with Consiglio Nazionale delle Ricerche, Istituto di Elettronica e di Ingegneria dell'Informazione e delle Telecomunicazioni, Italy (email: \{alberto.tarable, rudipaolo.paganelli, marcopietro.ferrari\}@cnr.it).   E. Storelli is with Exprivia SpA, Molfetta, Italy.  A. Gatto is with Politecnico di Milano, Dipartimento di Elettronica, Informazione e Bioingegneria, Milano, Italy (email:alberto.gatto@polimi.it). }


\maketitle
\begin{abstract}
This paper introduces Generalized Quantum-assisted Digital Signature (GQaDS), an improved version of a recently proposed scheme whose information theoretic security is inherited by adopting QKD keys for digital signature purposes. Its security against forging is computed considering a trial-and-error approach taken by the malicious forger and GQaDS parameters are optimized via an analytical approach balancing between forgery and repudiation probabilities.
The hash functions of the previous implementation are replaced with Carter-Wegman Message Authentication Codes (MACs), strengthening the scheme security and reducing the signature length. For particular scenarios where the second verifier has a safe reputation, a simplified version of GQaDS, namely deterministic GQaDS, can further reduce the required signature length, keeping the desired security strength.
\end{abstract}



\section{Introduction}
\label{sec:introduction}
Information technologies for remote data exchange are fundamental tools in modern society for managing relationships among people, legal entities or even machines. However, no matter how outstanding the technology development is, it would have been useless without cryptographic primitives able to protect our data from potential malicious entities trying to take advantage of confidential communication or dishonest personification. 

Confidentiality is managed by encryption and decryption algorithms using a secret key shared by the intended partners, usually denoted by Alice (the sender) and Bob (the recipient), whereas Eve generally denotes a malicious eavesdropper. 

Another powerful cryptographic primitive is Digital Signature (DS), which guarantees authenticity (the message originates from the claimed sender, Alice), integrity (the message is not altered by a forwarder, Bob) and non-repudiation (the sender cannot deny having sent the message because a verifier, e. g. Charlie, can recognize his signature on it).

The above sentences have a probabilistic meaning: the probability of dishonest events (fake, forgery, repudiation) can be kept as small as desired in an Information-Theoretically secure (IT-secure) scenario. At present, most available DSs are not IT-secure, having a security level based on the alleged computational difficulty of solving a class of problems for which, once the solution is provided, verification is manageably accomplished in an acceptable time and proves that the solution giver was the legitimate author of the riddle as well. 

The recent advent of Quantum Technologies provided the appealing ability to generate IT-secure keys and share them between two remote parties. IT-security directly follows from the laws of nature embedded in the Quantum Mechanics framework, which states the true randomness in the collapse of the wave function and the non-clonability of the physical system carrying information between two parties. The almost null chance for a third party (Eve) to remain unnoticed while gleaning information from photons travelling in the quantum channel makes the secret key shared by the two parties (Alice and Bob) IT-secure against violations: a process of this kind is called Quantum Key Distribution (QKD). 

On the other side, since passable scalable quantum computers may become soon available, Quantum Technologies represent a daunting threat as well, because Shor’s algorithm, taking advantage of quantum logic-gate operations, proved to be able to speed up the solution of discrete logarithm and factoring problems, on which present DS implementations built their computational security. For this reason, it is imperative to develop alternative DS methods, possibly relying on IT-secure implementations or, at least, post-quantum secure methods for which no faster algorithms have been devised (yet). 

A Quantum Digital Signature (QDS) was first proposed in 2001 in the seminal paper \cite{GottesmanChuang2001}, whose work can be thought of as an extension of a concept previously developed by Lamport in \cite{Lamport1979}, where he conceived a one-time signature scheme based on a one-way function; it is a function for which knowing the output (the image) gives no clue on which was the input (the pre-image). Among the merits of Lamport's one-time signature is that the signer (Alice) can generate the signature, destroy the key used in the process (the private key, the pre-image) and no longer interact with the verifiers: they only need the output of the one-way function (the public key, the image), anytime available from a trusted repository, and, with an ingenious method involving another one-way hashing function to firmly bond the particular message to the appropriate subset of keys to be verified, neither forgery nor repudiation is possible, making transferability feasible regardless of any inconvenience that may occur to Alice after the signature accomplishment, like loss of keys, theft or even death. 

The main idea in Gottesman and Chuang's work \cite{GottesmanChuang2001} was to use non-orthogonal states as "quantum one-way functions", which are guaranteed to be noninvertible by laws of quantum mechanics, even with unlimited computational power. The classical description of the state, which is only known to who prepares it (Alice), plays the role of the private secret key, whereas the quantum state itself plays the role of the public key with probabilistic test results, which cannot be tampered with without being spotted. 

The most striking feature of \cite{GottesmanChuang2001} is the development of the double-threshold verification method. Since the exchange of quantum states is a one-to-one process which involves the signer (Alice) and every possible verifier, a distribution phase is mandatory: all the verifiers, at least two, Bob (a first verifier with forwarding capability and possible commitment) and Charlie (a second verifier with impartiality and arbitrating intention) need to interact with Alice on a quantum channel to get the public key (a sequence of quantum states). Moreover, to confirm Alice provided them with the same sequence of quantum states, the verifiers need to interact in a one-to-one quantum channel with the so-called "swap test" that leaves the states undisturbed (if identical) or fails probabilistically (if different). Indeed, assuming a malicious Alice, the only means to confine her probability of success in repudiation is digging a trench between two thresholds and letting in almost all the repudiation attempts (probabilistically), as explained in the following. 

After the sign phase performed by Alice autonomously, anytime after the distribution phase, the verification phase will provide three possible outcomes in the comparison of the signature after-effect with the public key available to the verifier: 
\begin {itemize}
\item 1-ACC) it is almost certainly Alice's signature: the detected error rates are less than the hard threshold, whose value must be compatible with system non-idealities and background noise; 
\item 0-ACC) it is not confidently possible to exclude that it is Alice's signature, because of a sufficient degree of similarity: the detected error rate is more than the hard threshold but less than the soft one;
\item REJ) it is almost certainly not Alice's signature on the message: the detected error rate is more than the soft threshold, nonetheless, hard enough to cope with the best forger strategy attempts.
\end{itemize}

The DS implementation in \cite{GottesmanChuang2001} is affected by many practical restrictions: 1) the need for quantum computation capabilities for all the verifiers (to perform the swap test both in distribution and verification phases); 2) the need for a quantum memory to store the public key (the sequence of quantum states describing it); 3) the need for authenticated quantum channels between all the participants, 4) finally, the lack of "universal verifiability" because only parties that have participated to the distribution phase can verify the signature.

A first step toward more feasible QDS implementations has been proposed in \cite{Andersson2006}, where a replacement for the swap test has been introduced, which uses coherent states of light and realizable spatially separated multiport devices able to perform the needed quantum state comparisons, both in the distribution and verification phases. 

A second important step has been proposed in \cite{Dunjko2014}, where the need for a quantum memory has been circumvented by replacing the quantum public key with a classic verification key, which is no longer the same for all the verifiers, even in the ideal case, due to the random choice of the base used in the "unambiguous state discrimination" process that probabilistically extracts conclusive outcomes on some verification key elements (note that we no longer call them public keys, being different for Bob and Charlie). 

A third fundamental step has been proposed in \cite{Wallden2015}, where the replacement of the comparison test with a much more feasible "symmetrizing" exchange in the distribution phase leaves Alice without clues on how his private key information has been shared among the verifiers, and no means to fool one of them, but the random way, whose probability of success can be kept under control with the double-threshold method. The most significant feature of \cite{Wallden2015} is that the very same physical devices used for QKD are used to implement the QDS. Moreover, like in the QKD framework, the issue of authenticated quantum channels can be solved at the system level by "sacrificing" part of the communicated qubits on tamper tests and relying on pre-shared secrets between the intended parties. 

An experimental test of the scheme of \cite{Wallden2015} is described in \cite{Donaldson2016}, where the QDS protocol was implemented over 2-km links, and the transmission of the QDS for signing 1 bit was proven to be possible within a time window of 20 s, a huge improvement with respect to previous practical implementations. However, the length of the QDS is almost 2 billion qubits, and the security strength is reported to be 0.01\% (one  dishonest attempt of breaking the QDS protocol out of ten thousands succeeds, on average).

Following the guidelines described in \cite{Pirandola2020} for the implementation of a generic modern QDS protocol that does not require quantum memory, that can be realized with the same system devices used for QKD, and that does not require further assumption on the quantum channels used, reference \cite{CidetalArxiv23} proposes a new Quantum-assisted Digital Signature (QaDS). The QaDS protocol, which will be described in detail in the following, uses the QKD protocol to share IT-secure cryptographic keys between all pairs of agents in the QDS scenario, while the signature protocol itself is essentially classic, with extensive use of hash functions compliant with the most recent NIST recommendations. QaDS inherits the security of QKD while avoiding the main shortcomings of QDS, such as unfeasible long signatures and long delays needed to sign messages of arbitrary length. Since QKD is already a mature technology, QaDS promises to be an easily realizable surrogate of QDS, even if some of the limitations of QDS are not overcome by QaDS, such as the need for the signer to share a quantum link with all the conceivable verifiers.

In this paper, we propose Generalized QaDS (GQaDS), which is both a generalization and an improvement of QaDS, in many respects. In particular, we list below the main innovations we have introduced.
\begin{itemize}
    \item We introduce an extended analysis of the security against forging of the QaDS protocol, considering also a trial-and-error approach taken by Bob when attempting his forgery.
    \item We propose a semianalytical approach for the optimization of QaDS, where the sensitivity of the QaDS performance with respect to parameters such as the verification thresholds and the fraction of shared key blocks between Bob and Charlie is assessed and exploited in the design of GQaDS.
    \item We replace hash functions in the QaDS implementation with Carter-Wegman Message Authentication Codes (MACs), obtaining the double result of strengthening the scheme security and reducing the signature lengths.
    \item For particular scenarios where the second verifier (Charlie) is an authority endowed with a safe reputation, we propose a simplified version of GQaDS, named deterministic GQaDS, which further reduces the requirements in terms of signature length. 
\end{itemize}

In section II the QaDS proposed in \cite{CidetalArxiv23} is briefly summarized, and the main scheme parameters are introduced. In Section III, the security of QaDS is discussed, with a more advanced discussion on the security against forgery. Section IV deals with balancing the security against forgery and repudiation with several choices of protocol parameters, discussing their optimization under several possible constraints. Section V describes GQaDS, especially its novelties with respect to QaDS,  and deterministic GQaDS, which is introduced as an alternative scheme suitable for a particular scenario. Finally, conclusions are provided in section VI.  

\section{A Feasible Hybrid Quantum-assisted Digital Signature (QaDS)} \label{sec-QaDS from [7]}
In \cite{CidetalArxiv23}, the authors propose the minimal three-user scenario for a complete description of the procedure for signing and verifying variable-length messages using QKD keys. Alice is the signer, while Bob and Charlie are message verifiers playing different roles: for clarity, we will assume Bob is a first verifier (with forwarding commitment) using the hard threshold $V_B$, while Charlie is a second verifier (with arbitrating role) using the soft threshold $V_C$.

The protocol envisages two phases:
\begin{itemize}
    \item a \textit{distribution phase}, where Alice shares a QKD key with Bob and another one with Charlie, and where Bob and Charlie  share some portion of their respective QKD keys (Alice being unaware which portions are shared);
    \item a \textit{messaging phase}, where Alice signs the message and sends it to Bob, which either accepts or rejects the signature and, in the former case, forwards it to Charlie. 
\end{itemize}

In the first step of the distribution phase, Alice establishes symmetric QKD keys of $L$ bits with Bob and Charlie, denoted $k_{1}$ and $k_{2}$, respectively. 
During the second step of the distribution phase, Bob and Charlie share portions of their keys through a secure and authenticated channel (e.g., through the use of a pre-shared QKD key). In particular, Bob (Charlie) divides his key into $n$ blocks of length $r=L/n$, randomly selects half of these blocks and sends them to Charlie (Bob) together with their positions. 

The messaging phase starts with Alice signing an arbitrary-length message $m$, as follows. She applies the hash function $h(\cdot )$ on $m$, which outputs a digest $h_A=h(m)$ of fixed length $d=2L$. Next, Alice concatenates $k_1$ and $k_2$, obtaining a $d$-long local key  $k_A$ and uses it to encode $h_A$ by One-Time-Pad (OTP) encryption, obtaining $c_A = ENC_{k_A}(h_A)$. Then, Alice divides $c_A$ in $2n$ blocks, each of length $r$, and applies a hash function on each block, returning $2n$ digests of fixed $2L$ length. The $4nL$-long signature $S_A$ is generated by concatenating such digests, then Alice sends the tuple $(m, S_A)$ to Bob.

The next step of the messaging phase is Bob starting the verification procedure on the tuple $(m, S_A)$ received by Alice, which consists of performing the same operations executed by Alice for the generation of $S_A$.
Bob generates the new local key $k_B$ by concatenating the symmetric key shared with the signer ($k_1$), and the key blocks received from Charlie ($k'_2$) putting them in the correct position. He applies the same hash function $h(\cdot)$ as Alice on $m$, getting $h_B=h(m)$. Then, encoding $h_B$ with $k_B$ using OTP encryption, obtains $c_B=ENC_{k_B}(h_B)$ as a result. He splits $c_B$ into $2n$ blocks of length $r$ and applies the same hash function as Alice on each block. He gets the local signature $S_B$ by concatenating all the generated digests on $c_B$ blocks. Finally, Bob performs the validation process by comparing $S_B$ with $S_A$ using the threshold $V_B$. If the number of matching blocks surpasses $V_B$, Bob labels the message and the signature as authentic. Otherwise, he rejects the signed message as invalid.
If the signature is accepted, Bob forwards the tuple $(m, S_A)$ to Charlie to initiate the second verification procedure.

In the last step of the messaging phase, Charlie performs the same verification procedure as Bob, using the key blocks received from Bob ($k'_1$) and $k_2$  to generate his local key $k_C$ by concatenation. He calculates the hash of the message $h_C=h(m)$ and encrypts it using OTP with $k_C$, getting $c_C=ENC_{k_C}(h_C)$. Then, he splits $c_C$ into $2n$ blocks of length $r$ and applies the hash function on each block, obtaining the sequence of digests. He generates the local signature $S_C$ by concatenating all the digests and compares $S_C$ with $S_A$ using the verification threshold $V_C$. Like Bob, if the number of matching blocks surpasses $V_C$, Charlie accepts the signed message; otherwise, he rejects it as invalid. 

Table \ref{tab1} summarises a list of parameters used in the process.

\begin{table}
\caption{List of parameters involved in the (G)QaDS process}
\label{table}
\setlength{\tabcolsep}{2pt}
\centering
\begin{tabular}{|p{25pt}|p{250pt}|}
\hline
Symbol&  
Description \\
\hline
$L$& 
QKD key length\\
$n$& 
Number of blocks in a QKD key\\
$r$& 
QKD Key block length (r=$L/n$)\\
$ m $& 
Arbitrary-length message to sign\\
$ k_{1} $& 
QKD key between Alice and Bob \\
$k_{2}$& 
QKD key between Alice and Charlie\\
$k_{A}$& 
Concatenated QKD key that Alice uses to sign the message\\
$ k'_{1} $&
Random key blocks disclosed from $k_1$ \\
$ k'_{2} $&
Random key blocks disclosed from $k_2$\\
$ h_A $& 
Hash of the message \textit{m} calculated by Alice \\
$ h_B $& 
Hash of the message \textit{m} calculated by Bob \\
$ h_C $& 
Hash of the message \textit{m} calculated by Charlie \\
$ c_A $& 
OTP string generated by Alice using $k_A$ and $h_A$ \\
$ c_B $& 
OTP string generated by Bob using $k_b$ and $h_b$ \\
$ c_C $& 
OTP string generated by Charlie using $k_c$ and $h_c$ \\
$ S_A $& 
Alice's signature on the message \textit{m}\\
$ S_B $& 
Bob's local signature on the message \textit{m} with $k_1$ and $k'_2$ as input\\
$ S_C $& 
Charlie's local signature on message \textit{m} with $k'_1$ and $k_2$ as input\\
$ S_f $& 
Alice's signature forged by Bob and forwarded to Charlie\\
$ V_B $& 
Bob's verification threshold (first level verifier) \\
$ V_C $& 
Charlie's verification threshold (second level verifier) \\
\hline
\end{tabular}
\label{tab1}
\end{table}

\section{QaDS security evaluations}

Authors in \cite{CidetalArxiv23} take for granted that exactly one-half of the key blocks have to be shared among the verifiers, to cope with repudiation attempts and forgery attempts in a balanced way. Moreover, it is understood that Bob's hard threshold implies zero errors must be spotted on any of the $n+n/2$ blocks known to Bob (i.e. $V_B=(3/2)n$ for 1-ACC verification): errors are only allowed to happen (and, in a fair scenario, almost certainly do) on the $n/2$ blocks Charlie kept reserved. 
Now, we can briefly illustrate the security proofs proposed in \cite{CidetalArxiv23} and spot some novel contributions.

\subsection{Repudiation proof}
In this case, Alice is the malicious user who wants to generate a signature that Bob accepts but Charlie rejects, thus successfully denying the authorship of her message. Therefore, Alice must introduce as many errors in the blocks of $k_2$, on which Bob has no information, as necessary to violate the soft threshold $V_C$. 
The higher the error number compatible with the soft threshold $V_C$, the higher the number of errors Alice needs to introduce, and the higher the probability that at least one error ends on a block that Charlie disclosed to Bob during the distribution phase, making Bob aware of the repudiation attempt.

In particular, we want to evaluate the probability that none of the tampered blocks fall into the set Charlie shared with Bob. Therefore, the probability that Alice succeeds in an attack of repudiation is calculated as follows:
\begin{equation}P_{R}=\prod_{i=0}^{e-1}\frac{n/2-i}{n-i} \label{eq2}\end{equation}
where $e$ represents the number of errors introduced by Alice.

\subsection{Integrity and Hash function features}
In this scenario, Bob is the malicious user who wants to substitute the message $m$ with a different message $M\neq m$ and to make Charlie's verification process succeed. Since the calculation of $S_i$ depends on the value of $h_i$, the alteration produced by the message modification is propagated along the entire chain, triggering an error in almost any block of Charlie's verification test. Assuming that Bob recalculates Alice's signature (for $M$) in the blocks for which he knows the private key, Bob attack on integrity is successful by finding a second partial preimage $M \neq m$ that provides the same hash value of the original message in the positions of a sufficient number of Charlie's private blocks compatible with the $V_C$ threshold satisfaction, and by filling those blocks with the true Alice's $S_A$ values, meaning that the robustness against integrity involves the hash function properties.

\subsection{Forgery proof and key block guessing}
A forgery attack consists in Bob forging Alice's signature. Specifically, when he receives the tuple $(m, S_A)$ from Alice, he generates a tampered message $M \neq m$ associated with a forged signature $S_f \neq S_A$ and sends it to Charlie. The attack is successful if Charlie accepts $(M, S_f)$ as valid after the verification procedure.

Authors in \cite{CidetalArxiv23} propose three possible ways for Bob to forge $S_f$: the first two are based respectively on finding a preimage of $m$ and on extracting the unknown elements of $k_2$ from the signature $S_A$, both of which have been considered in the previous subsection; the third is based on the probability of guessing the unknown bits of $k_2$. In particular, the probability of success in a forgery attack was coarsely computed in \cite{CidetalArxiv23} as: 
\begin{equation}
P_{F}=\frac{1}{2^{L/2}}, 
\label{eq1}\end{equation}
which is rather optimistic, for two reasons. 

Even finding a single block of Charlie's private key can work if the $V_C$ threshold is tuned for only one block to comply with the 0-ACC verification (i.e. $V_C=n+1$): in this case we should use $L/n$ instead of $L/2$ in \eqref{eq1}. Anyway, $P_F$ must depend on $V_C$. 
Moreover, the security analysis against forgery attempts shown in \cite{CidetalArxiv23} is based on the tacit hypothesis that Bob is given a one-shot attempt to guess a number of Charlie's private key blocks larger than $V_C$ (Charlie's threshold for rejection). However, Bob has the chance to check the correctness of his guesses by applying the candidate key blocks to the true message $m$ and comparing the results with the signature blocks corresponding to Charlie's private key, as available from Alice's true signature $S_A$. Thus, Bob can implement a trial-and-error strategy based on an exhaustive search of all possible key-block values, until a sufficient number of Charlie's private blocks is found. 

In our novel analysis, Bob starts checking each of the $2^{r}$ possible values of the key block, until he finds $V_C-n$ correct key blocks (the minimum possible to make Charlie's verification succeed). Let us define the computation complexity of Bob's forging as the average number of block values he has to test. The following proposition holds.

\begin{proposition} \label{prop:forging}
The computation complexity of Bob's forging is
\beq \label{eq:CF}
\Cc_F = \left(V_C-n\right) \frac{2^{r}+1}{n/2+1}.
\eeq
\end{proposition}

\pf 
The problem is equivalent to the following one, with $M = 2^r$, $K = n/2$, $k = V_C-n$.

\emph{Problem: Consider an urn with $M$ balls, out of which $K$ are green and the remaining red.  Find the average number of picks without replacement to get $k$ green balls.}

Let $V$ be the random variable representing the number of picks to be performed to get $k$ green balls.  $V = v$ happens if $k-1$ green balls are collected in the first $v-1$ picks,    and the $v$-th pick yields the $k$-th green ball. Then, We have:
\begin{eqnarray}
\PP\{V = v\} &=& \frac{\binom{K}{k-1} \binom{M-K }{ v-k}}{\binom{M }{ v-1}} \frac{K-k+1}{M-v+1}\non
&=& \frac{\binom{K-1}{k-1} \binom{M-K }{ v-k}}{\binom{M -1}{ v-1}} \frac{K}{M}=  \frac{\binom{v-1}{k-1} \binom{M-v }{ K-k}}{\binom{M -1}{ K-1}} \frac{K}{M}\non
&=& \frac{\binom{v-1}{k-1} \binom{M-v }{ K-k}}{\binom{M }{ K}}
\end{eqnarray}

Thus, the average value of $V$ will be given by
\begin{eqnarray}
E[V] &=& \sum_{v=1}^M v \PP\{V = v\} \non
& = & \frac1{\binom{M }{ K}} \sum_{v=1}^M v \binom{v-1}{k-1} \binom{M-v }{ K-k} \non
& = & \frac{k}{\binom{M }{ K}}  \sum_{v=1}^M \binom{v}{k} \binom{M-v }{ K-k} \non
& = & \frac{k}{\binom{M }{ K}} \binom{M+1 }{ K+1} = k \frac{M+1}{K+1}
\end{eqnarray}

The distribution of $V$ is strictly related to the negative hypergeometric distribution.
\qedsymb


\section{Balancing security against forgery and repudiation} \label{sec-P_F P_R optimization}

In \cite{CidetalArxiv23}, many parameters, such as the verification thresholds, the number of deliberate errors injected by a malicious Alice for a repudiation attempt, and the number $S$ of key blocks sent from Bob to Charlie (or \textit{vice versa}), are left as understood. For instance, $S$ is assumed by default as $n/2$, corresponding to half of the key length. Intuitively, the largest is $S$, the easiest is a success in a forgery attempt by a malicious Bob. Conversely, the hardest is for a malicious Alice to confine her deliberate errors in  Charlie's private key. Hence an intermediate value $S=n/2$ may seem an optimized choice. However, these probabilities also depend on the thresholds $V_B$ and $V_C$, and they are different functions of $S$, thereby leading to a nontrivial optimization. In this section, we revise the matter and discuss appropriate values for each parameter.

\subsection{Repudiation probability}

Let $V_B$ and $V_C$ represent the verification thresholds, i.e., the minimum number of correct signature blocks requested by Bob and Charlie respectively to accept the validity of Alice's signature. Thus, Bob (Charlie) rejects Alice's signature if the number of mismatched blocks $e_B (e_C)$ among the $n+S$ he can check\footnote{Note that Bob must validate Alice's signature by checking \underline{all} the blocks he knows including the $S$ received by Charlie, otherwise Alice can easily repudiate.} is larger than $n+S-V_B$ ($n+S-V_C$). Alice succeeds in repudiation by corrupting $e$ signature blocks if and only if $e_B \leq n+S-V_B$ and $e_C > n+S-V_C$, thus

\begin{equation}
\PP\{Rep.|e\}=\PP\{e_B \leq n+S-V_B \cap e_C > n+S-V_C |e\}. \nonumber
\end{equation}

In our scenario without channel errors (the exchange of signature blocks is protected by errors at data-link layer), it is reasonable to set $V_B = n+S$, meaning that Bob has no tolerance on errors, since any mismatched block observed by Bob is deliberate and reveals a repudiation attempt by Alice. On the other hand, Alice must inject $e>n+S-V_C$ mismatches, otherwise $\PP\{Rep.|e\}=0$. Besides, Alice must concentrate the mismatches in Charlie's blocks, and repudiation succeeds if and only if none of these errors occurs in the $S$ blocks that Charlie sent to Bob during the distribution phase, as these are checked by Bob as well.

Assuming a random selection, unknown to Alice, of the $S$ key blocks revealed by Charlie to Bob, the pmf (given $e$) of the number of mismatched blocks observed by Bob $e_B$ obeys a hypergeometric law of parameters $(n,n-S,e)$, i.e.  

\begin{equation}
    \PP\{e_B=x|e\}=\mathcal{H}_{n,n-S,e}(x)=\frac{\binom{n-S}{e}\binom{S}{x}}{\binom{n}{e}}
\end{equation}
since no mismatched block can be found by Bob among his blocks, if Alice concentrates her errors in Charlie's ones. Therefore

\begin{equation}
    \PP\{Rep.|e\}=\mathcal{H}_{n,n-S,e}(0)=\prod_{i=0}^{e-1}\frac{n-S-i}{n-i}
    \label{eq - P[Rep|e]}
\end{equation}
which is the generalization of \eqref{eq2} with an arbitrary number $S$ of key blocks exchanged between Bob and Charlie. 

Since all factors in the RHS of \eqref{eq - P[Rep|e]} are smaller than one, $\PP\{Rep.|e\}$ decreases with increasing $e$. Therefore, the best strategy for Alice is to inject the minimum necessary number of errors, i.e., 
\begin{equation}
 e^*=n+S-V_C+1   \label{eq-e^star}
\end{equation}
assuming that Alice knows $V_C$, which is a conservative assumption since she can try to guess it. Hereafter, we assume an optimized repudiation strategy that succeeds with probability

\begin{equation} \label{eq:PR0}
    P_R=\PP\{Rep.|e^*\}=\prod_{i=0}^{n+S-V_C}\frac{n-S-i}{n-i}
\end{equation}

Now, we optimize the parameters of GQaDS by means of a semianalytical approach. In the following,  we set $S = \beta n$, where $\beta$ ($0 < \beta < 1$) is  a constant, which is equal to 1/2 for QaDS. Regarding $V_C$, we set
\beq
V_C = 2 \beta n + \gamma (1-\beta) n
\eeq
where $\gamma$ ($0 \leq \gamma \leq 1$), which may depend on $n$, is the fraction of Charlie's private blocks that need to be correct for Charlie to accept Alice's signature.  Notice that $\gamma = 0$ means that Charlie does not care about the secret part of his key (and thus repudiation is impossible), while $\gamma = 1$ means that Charlie has zero tolerance on errors (and thus repudiation succeeds with probability $1-\beta$).
Substituting this parameterization in \eqref{eq:PR0} and making some simplifications, we obtain 
\beq \label{eq:P_rep_estar}
P_R = \frac{((1-\beta)n)!}{n!}	\frac{((\beta + \gamma(1-\beta))n - 1)!}{( \gamma(1-\beta)n - 1)!} .
\eeq

\begin{figure}[!ht]
\centering
\centerline{\includegraphics[width=0.6\textwidth]{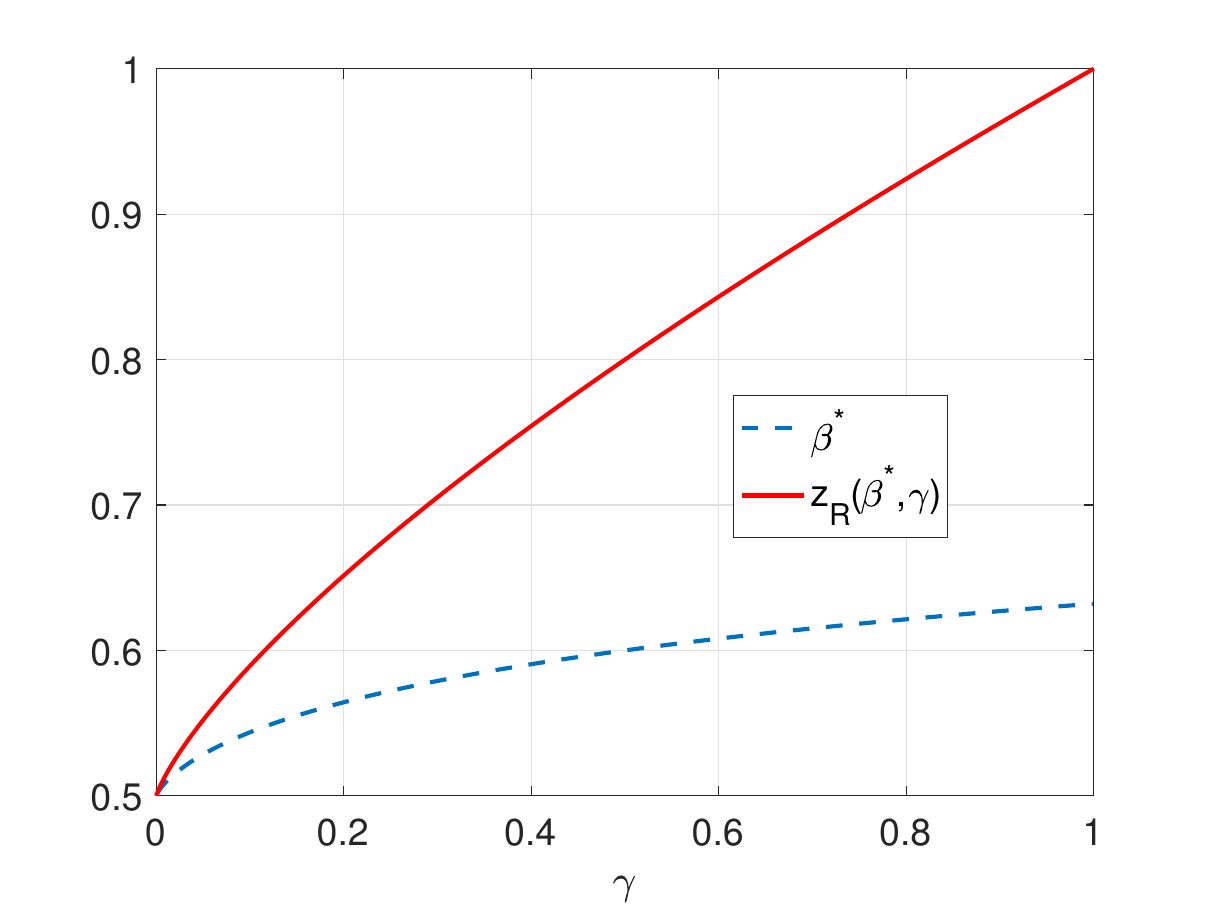}}
\caption{$\beta^*(\gamma)$ and $z_R\left(\beta^*(\gamma),\gamma\right)$ according to \eqref{eq:beta_star} and \eqref{eq:zeta_star}.}
\label{fig:beta_vs_gamma}
\end{figure}

Now we distinguish two different scenarios.
\begin{itemize}
    \item In the first, $\gamma (1-\beta) n$ tends for sufficiently large $n$ to a constant $C$, which happens if $\gamma = O(1/n)$. In this case, for large $n$, we approximate \eqref{eq:P_rep_estar} as
    \begin{eqnarray}
        P_R &\approx&  \frac{((1-\beta)n)!}{n!}	\frac{(\beta n + C-1)!}{(C-1)!} \nonumber  \\
        &\approx& \left({n \choose \beta n}(C-1)! \right)^{-1} \nonumber
        \\
        &\approx & 2^{-nh(\beta)} / (C-1)! \label{eq:P_rep_app1}
    \end{eqnarray} 
    where $h(\beta) = - \beta \log_2 \beta - (1-\beta) \log_2 (1-\beta)$ is the binary entropy function. The last approximation involves Stirling's approximation of the binomial coefficient. From \eqref{eq:P_rep_app1}, we see that, in this scenario, $\beta^* = 1/2$, as for QaDS, is the value that minimizes $P_R$.

    \item In the second, $\gamma (1-\beta) n$ grows linearly with $n$ for large $n$. In such regime, we can use Stirling's approximation for the factorials in \eqref{eq:P_rep_estar}, thus obtaining 
    \beq \label{eq:P_rep_app}
    P_R \approx z_R(\beta,\gamma)^n \frac{\gamma(1-\beta)}{\beta + \gamma(1-\beta)}
    \eeq
    where
    \beq
    z_R(\beta,\gamma) = \frac{(1-\beta)^{1-\beta} (\beta + \gamma(1-\beta))^{\beta + \gamma(1-\beta)}}{(\gamma(1-\beta))^{\gamma(1-\beta)}}.
    \eeq
    Thus, for large $n$, the right-hand side of \eqref{eq:P_rep_app} is minimized if  $z_R(\beta,\gamma)$ is the smallest possible. A straightforward computation yields 
    \beq \label{eq:beta_star}
    \beta^*(\gamma) = \frac{\left(1/\gamma\right)^{\frac{\gamma}{1-\gamma}} - \gamma}{\left(1/\gamma\right)^{\frac{\gamma}{1-\gamma}} - \gamma + 1}
    \eeq
    as the optimal value of $\beta$, i.e., the one that minimizes $z_R(\beta,\gamma)$. Correspondingly, we easily obtain
    \beq \label{eq:zeta_star}
    z_R(\beta^*(\gamma),\gamma) = \frac{\left(1/\gamma\right)^{\frac{\gamma}{1-\gamma}}}{\left(1/\gamma\right)^{\frac{\gamma}{1-\gamma}} - \gamma + 1}.
    \eeq
    Figure \ref{fig:beta_vs_gamma} shows the behavior of $\beta^*(\gamma)$ and of $z_R\left(\beta^*(\gamma),\gamma\right)$ as a function of $\gamma$. They are both increasing, convex functions, starting from $\beta^*(0) = z_R\left(1/2,0\right) = 1/2$ and reaching $\beta^*(1) = 1-1/\ee$ and $z_R\left(\beta^*(1),1\right) = 1$, respectively.

\end{itemize}

\subsection{Forging probability}

Bob succeeds in forging Alice's signature if he guesses the correct value of at least $\gamma(1-\beta)n$ blocks unknown to him, thus inducing Charlie to accept the false signature. Let $p = 2^{-r}$ be the probability of guessing correctly one entire key block. The average number of attempts he must perform in order to recover the due number of blocks will be given by
\beq
\Cc_F(\beta,\gamma) = \gamma(1-\beta)n \frac{1+1/p}{1+(1-\beta)n} \label{eq-C_F}
\eeq
which is the extension of \eqref{eq:CF} to our generalized parametrization. We define the forging probability as
\beq \label{eq:PF}
P_F = \Cc_F(\beta^*(\gamma), \gamma)^{-1}.
\eeq
Like for repudiation, we distinguish two scenarios, according to the behaviour of $V_C$ with $n$. If $\lim_{n \rightarrow \infty} \gamma (1-\beta)n  = C$, we have, for large $n$ and small $p$
\beq
P_F \approx \min \left\{1, \frac{(1-\beta)np}{C} \right\}
\eeq
which tends to 1, as $n \rightarrow \infty$, unless $p = O(1/n)$. If instead $\gamma(1-\beta)n$ gets large for large $n$, $P_F$ can be written for large $n$ and small $p$
\beq
P_F \approx \min \left\{1, \frac{p}{\gamma} \right\}
\eeq
independently of $\beta$.

\subsection{Optimization of $\gamma$ with fixed $n$ and $r$}

In this subsection, we show numerical results, giving the optimal value of $\gamma$, defined as
\beq \label{eq:PF_app}
\gamma^* = \arg \min_{\gamma \in [0,1]} \max\{P_R, P_F \}.
\eeq
Since $P_R$ is increasing with $\gamma$, as can be easily seen numerically, and $P_F$ is decreasing with $\gamma$, the minimum is obtained for $P_R = P_F$, if this condition is satisfied by some value of $\gamma$, as is often the case.   

\begin{figure}[!ht]
\centering
\centerline{
\includegraphics[width=0.6\columnwidth]{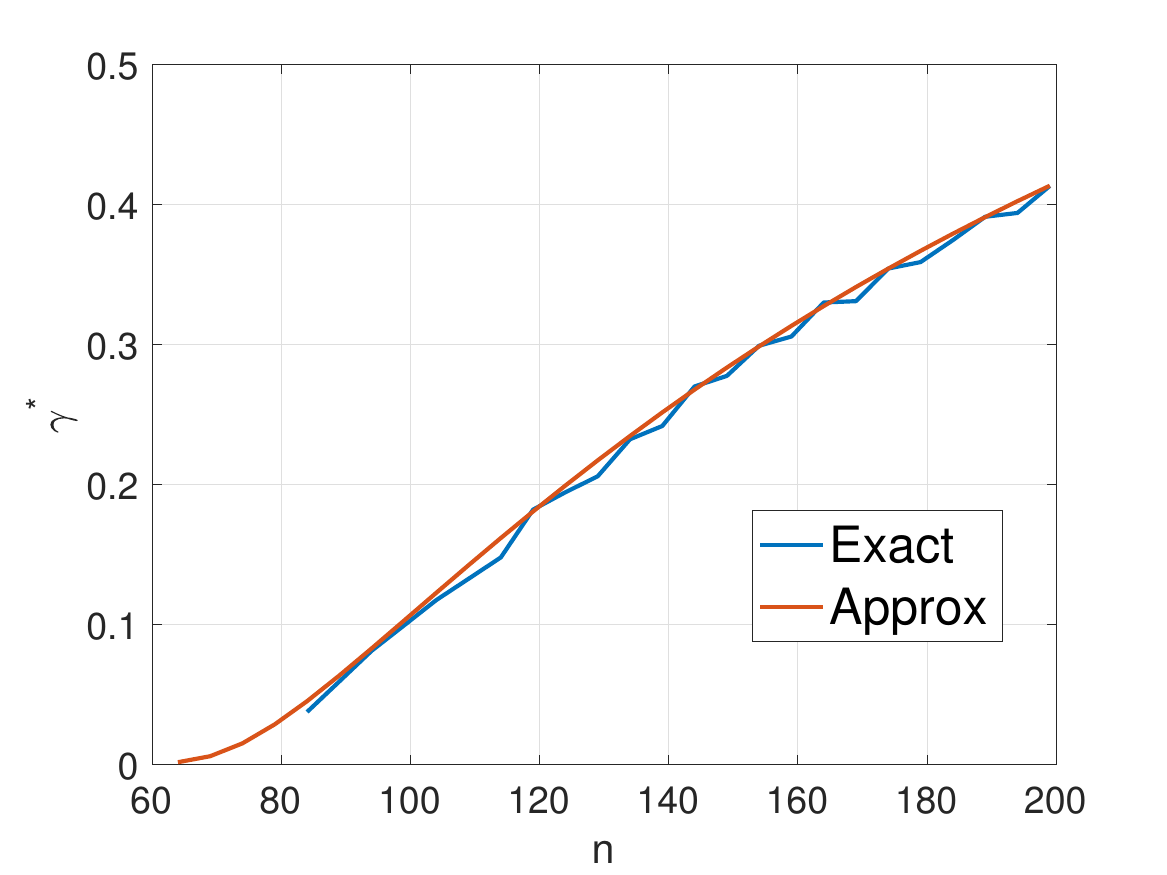}}
\caption{The optimal value of $\gamma$ as a function of $n$, $\gamma^*$ as defined in \eqref{eq:PF_app}, for $r = 82$.}
\label{fig:gamma_vs_N}
\end{figure}

\begin{figure}[!ht]
\centering
\centerline{\includegraphics[width=0.6\columnwidth]{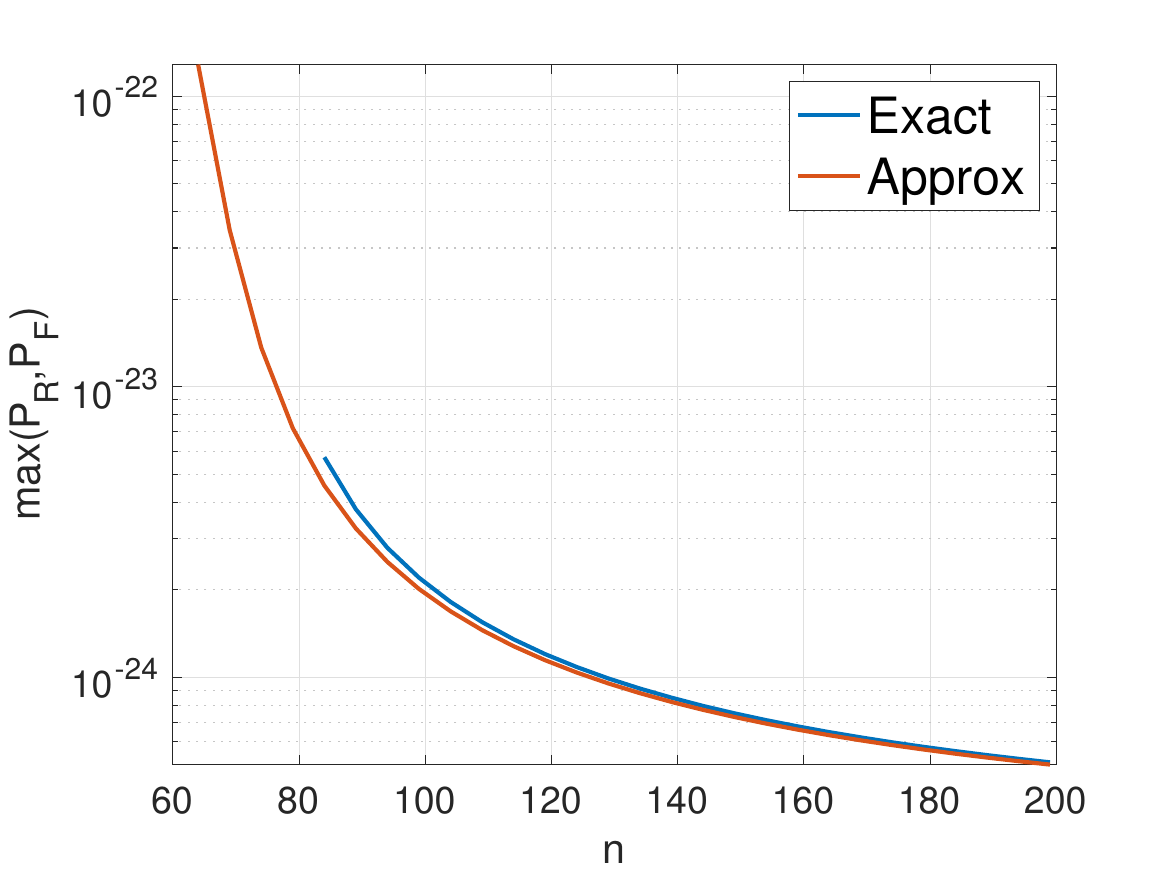}}
\caption{The lowest achieved value of $\max\{P_R, P_F\}$ as a function of $n$, for $r = 82$.}
\label{fig:PR_vs_N}
\end{figure}

In Fig.~\ref{fig:gamma_vs_N}, we show $\gamma^*$ as a function of $n$ for  $r=82$. There are two curves, both of which are obtained numerically. The first,  with label 'Exact', is obtained by considering the \textit{exact}  expressions of $P_F$ and $P_R$ in \eqref{eq:P_rep_estar} and \eqref{eq:PF}, while the second, labelled 'Approx', shows the value of $\gamma^*$ when the approximate expressions for $P_F$ and $P_R$ in \eqref{eq:P_rep_app} and \eqref{eq:PF_app} are computed.  The differences are mostly due to rounding effects. The corresponding values of $\max\{P_R,P_F\}$ is shown in Fig.~\ref{fig:PR_vs_N}. As it is reasonable, increasing $n$ improves the system robustness to malicious attacks.

\subsection{Optimization of GQaDS parameters with fixed key length}

In this section, we show results that allow to understand what is the optimal choice of $n$ given a fixed key length $L$. The results we show are for $L \in \{L_i\}_{i=0}^{11}$, where $L_i = 256 \times 2^i$. For a given value of $L$, let $n_{\mathrm{opt}}$ and $r_{\mathrm{opt}} = L / n_{\mathrm{opt}}$ be the optimal number of blocks and key block length, respectively, where
\beq \label{eq:nopt}
n_{\mathrm{opt}} = \arg \min_{n | L} \max\{P_R, P_F \}
\eeq
so that the optimization for a given $n$ follows the same line of the previous subsection.

\begin{figure}[!ht]
\centering
\centerline{\includegraphics[width=0.6\columnwidth]{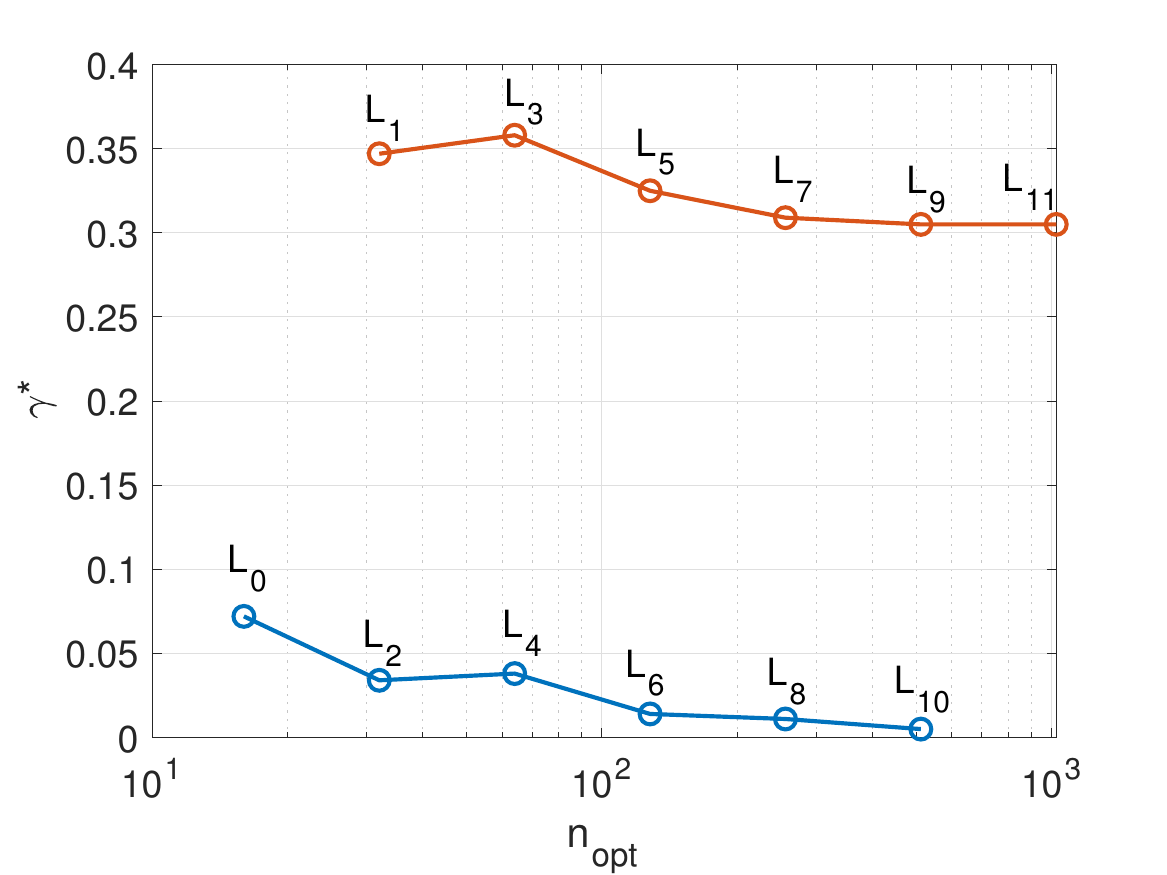}}
\caption{$\gamma^*$ versus $n_{\mathrm{opt}}$ for the set of key lengths $\{L_i\}_{i=1}^{11}$.}
\label{fig:gamma_vs_N_l_fixed}
\end{figure}

\begin{figure}[!ht]
\centering
\centerline{\includegraphics[width=0.6\columnwidth]{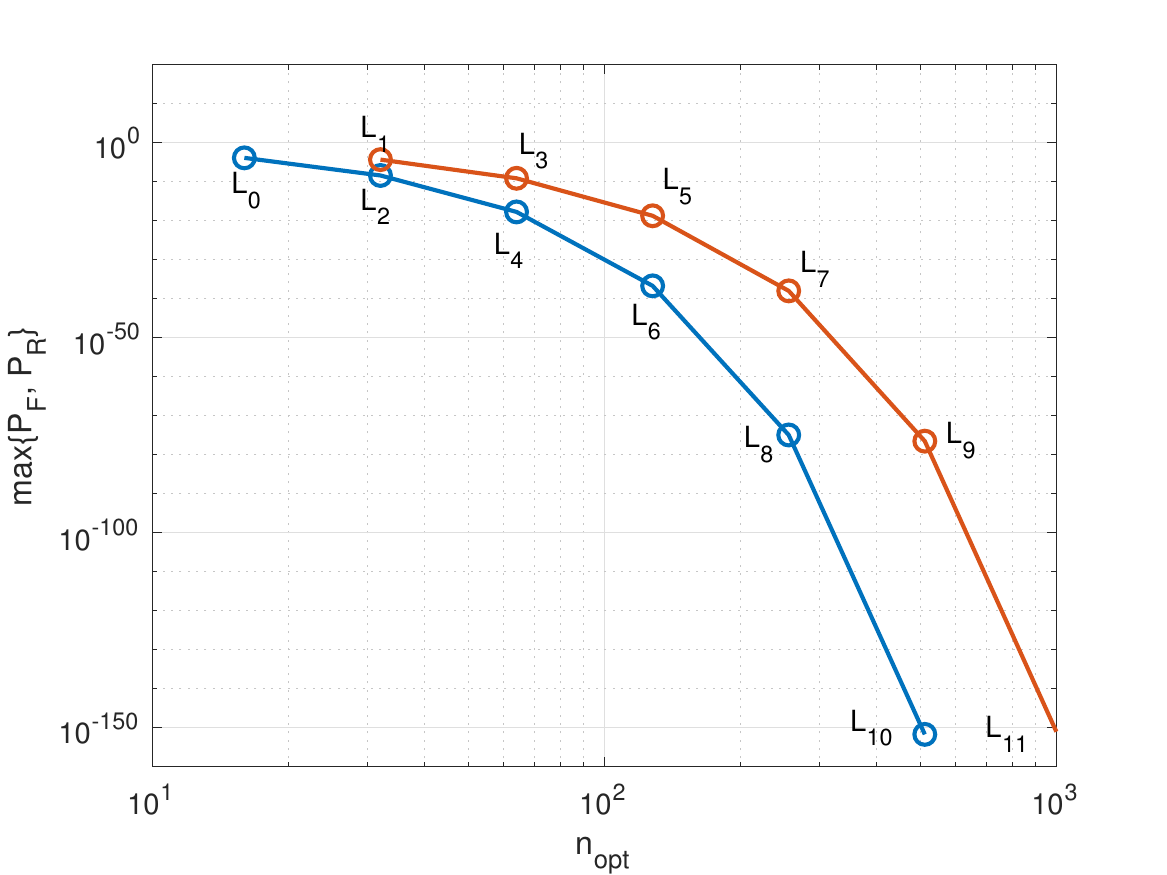}}
\caption{$\max\{P_F,P_R\}$ versus $n_{\mathrm{opt}}$ for the set of key lengths $\{L_i\}_{i=1}^{11}$.}
\label{fig:P_opt_vs_N_l_fixed}
\end{figure}

Figure \ref{fig:gamma_vs_N_l_fixed} shows the value of $\gamma^*$  while Figure \ref{fig:P_opt_vs_N_l_fixed} shows the corresponding value of $\max\{P_F,P_R\}$, both as a function of $n_{\mathrm{opt}}$. From the two figures, we can make the following observations.
\begin{itemize}
\item When $L = L_i$ with $i$ even, it is a perfect square and the optimal values are given by $n_{\mathrm{opt}} = r_{\mathrm{opt}} = \sqrt{L}$. In such a case, $\gamma^*$ is small and tends to zero for increasing $n_{\mathrm{opt}}$, so that $\beta^* \approx 1/2$, in accordance with \eqref{eq:P_rep_app1}. Notice that for small $\gamma$ and $n = r$, $P_R \approx 2^{-n}$ and $P_F \approx n 2^{-n}$, so the asymptotic behavior is about the same. As the optimization should force $P_F \approx P_R$, the results are in line with the theory.

\item When $L = L_i$ with $i$ odd, instead, $L$ is not anymore a perfect square, and $n_{\mathrm{opt}} = r_{\mathrm{opt}} / 2.$\footnote{The relationship is not true in general for any $L$, but heuristically, we have observed that $n_{\mathrm{opt}}$ and $r_{\mathrm{opt}}$ are still as balanced as possible.} Because of this asymmetry, $\gamma^*$ does not tend to 0 for increasing $n_{\mathrm{opt}}$, and $\beta^* = 1/2$ is no longer optimal. In this particular case, $\gamma^* \approx 0.3$ and $\beta^* \approx 0.58$. It is however to be noted that the minimum forging and repudiation probabilities for $L_{2i+1}$ are the same as for $L_{2i}$, so this configuration is suboptimal, as doubling the key length does not improve the scheme performance.

\end{itemize}

From the previous observations, we can conclude that the best values of $L$ are those that are (close to) perfect squares. In such cases, it is convenient to choose $n \approx r \approx \sqrt{L}$, $\beta \approx 1/2$ and $\gamma$ small (or, more precisely, $\gamma (1-\beta) n$ a small integer). The value of $L$ should be chosen large enough to obtain a target value of $P_F$ and $P_R$. If instead $L$ is constrained to be a predefined value, the above analysis is able to predict the optimal values of the parameters.  

\subsection{Asymmetric constraints on forgery and repudiation}

It is reasonable to assume that the constraints put on repudiation and forgery probabilities are not symmetric. Indeed, while forgery can be seen as a long-term attack, with a consequently large time horizon available for a dishonest Bob to make his forgery attempt successful, Alice's repudiation is a one-shot attack, whose probability of success can then be considered as less critical for the scheme to be practically acceptable. 

Let us suppose that values $P_R < 2^{-b_R}$ and $P_F < 2^{-b_F}$ are acceptable, where $b_R$ and $b_F$, $b_R < b_F$, are minimum security strengths (in bits) for repudiation and forgery, respectively. Let also $\alpha = 2^{b_F-b_R} > 1$ be an asymmetry coefficient. We can modify the optimization in \eqref{eq:nopt} as 
\beq
n_{\mathrm{opt}}^{\mathrm{asym}} = \arg \min_{n | L} \max\{P_R, \alpha P_F \}
\eeq
which is tantamount to say that we consider an auxiliary key block length $r' = r - b_F + b_R$. From the previous subsection, we obtain that, given a sufficiently large $L'$ needed to meet the requirements, the optimal parameters are $n \approx r' \approx \sqrt{L'}$, $\gamma$ small and $\beta \approx 1/2$. Then, the actual key length will be given by $L = n r = L' + n(b_F-b_R)$.

\section{GQaDS }

\subsection{{Parameter definition}}

In \cite{CidetalArxiv23},  practical secure values for robustness to forgery and repudiation are not considered. In the last years, signatures with security strength below 80 bits (i.e., $P_R , P_F > 2^{-80}$) are considered no longer secure, and for long-term security a strength of at least 112 bits is recommended \cite{NIST1,NIST2}. In our GQaDS scenario, long-term security is reasonably required against forgery attacks, whereas security against repudiation attacks that are one-shot can be still considered acceptable with a strength of 80 bits. Therefore we assume as target $P_R<10^{-24}$ and $P_F<10^{-40}$.

\begin{figure}[t]
\centering
\centerline{\includegraphics[width=0.6\textwidth]{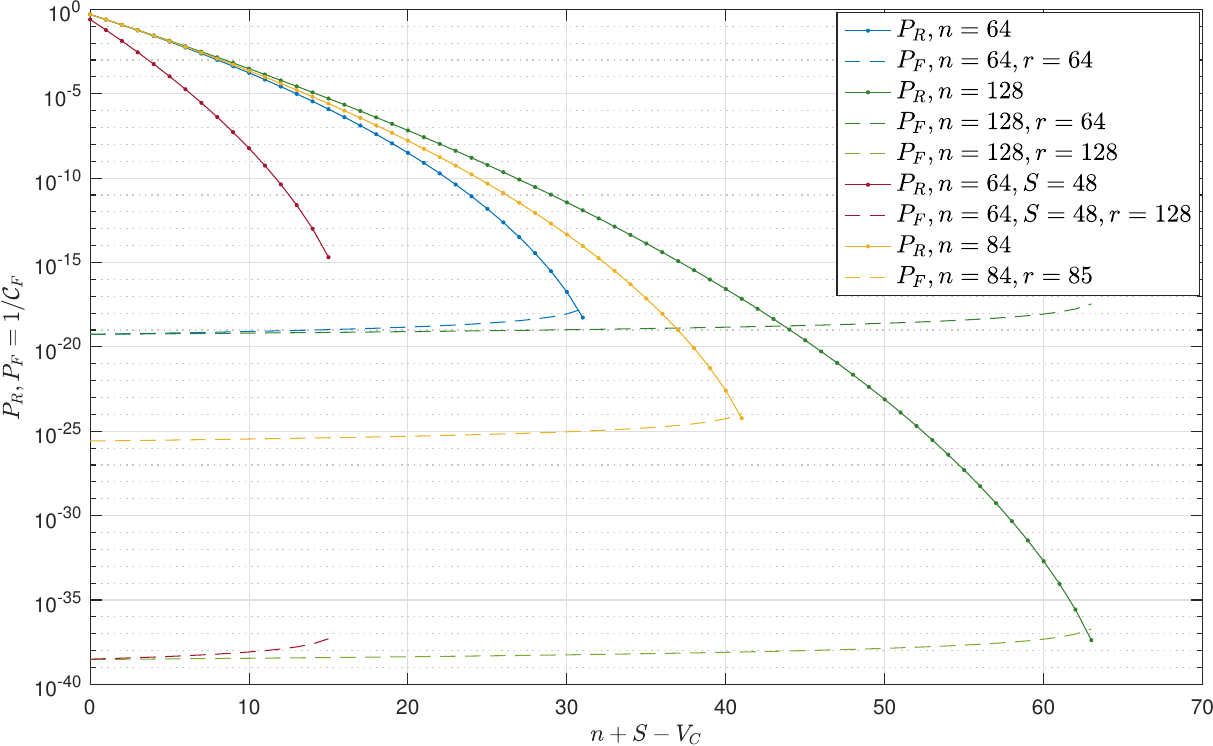}}
\caption{Probability of repudiation and forgery as functions of $n+S-V_C$ for various $n$ and $r$.}
\label{fig:Prep Pfrg vs Tc}
\end{figure}

As apparent from \eqref{eq:PR0} and \eqref{eq-C_F}-\eqref{eq:PF}, $P_R$ depends on $n,S$ and $V_C$, whereas $P_F$ depends mostly on $r$. In Fig. \ref{fig:Prep Pfrg vs Tc} we plot $P_F$ and $P_R$ as functions of $n+S-V_C$ for various $n, S$ and $r$. For any curve the abscissa is limited by $n+S-V_C<n-S$, otherwise $V_C \leq 2S$ and the probability of forgery $P_F \rightarrow 1$ because $\gamma=0$ in \eqref{eq-C_F}. Note that $P_R$ decreases with increasing $n+S-V_C$, whereas $P_F$ is almost constant as it mainly depends on $r$. In fact, the red and green dashed curves that share the same value of $r=128$ offer the same security strength against forgery (approximately 125 bits) with different number of blocks $n$. Also note that in the case $n=64,S=48$, the minimum achievable $P_R$ is $10^{-15}$ with $n+S-V_C=15$ that corresponds to the minimum possible threshold $V_C=2S+1=97$. In fact, even if increasing $S$ reduces the number of blocks in the private key of Charlie, Alice's strategy also reduces the number of errors inserted in the signature, increasing the minimum $P_R$ that GQaDS can guarantee. 
From the curves in Fig. \ref{fig:Prep Pfrg vs Tc} we see that in order to guarantee $P_F=P_R \approx 10^{-18}$, both $n=64$ and $r=64$ are needed at least, and up to $n=84$ and $r=85$ to reach $P_F=P_R \approx 10^{-24}$.

\subsection{Carter-Wegman MAC}

In QaDS,  after producing the OTP $c_A=ENC_{k_A}(h_A)=h_A \bigoplus k_A$, Alice divides $c_A$ into $2n$ blocks and applies individual hash functions to each block getting the final signature $S_A$, of length $|S_A|=4rn^2$.
With $n=r=64$ the signature size is $|S_A|=2^{20}$.    This blockwise hash is necessary, since otherwise, from $c_A$ and $h_A$, Bob could derive $k_A$ and hence $k_2$  by XOR-ing $h_A$ and $c_A$.

The signature length can be reduced by using a Carter-Wegman MAC instead of a hash function.  On a given message $m$, the Carter-Wegman MAC algorithm first obtains $h(m)$ as the output of a universal hash function (UHF), which can be implemented by interpreting $m$ as the coefficient vector of a polynomial on a fixed finite field and evaluating the polynomial at a given secret point $s$. Then, the UHF output $h(m)$ is input to a pseudo-random function (PRF) $F$, with secret key $k$, to obtain the MAC $c = \mathrm{MAC}(m,k) = F(k, h(m))$. With a Carter-Wegman MAC, it is not possible to obtain $k$ from $c$ and $m$, unless by brute force, because of the security of the PRF and the collision resistance of the UHF.

In GQaDS, Alice divided $k_A$ into $2n$ blocks $k_{A,i}, i=1,2...2n$, and builds the final signature $S_A$ as the concatenation of $2n$ MACs $c_i = \mathrm{MAC}(m,k_{A,i})$, for a signature length $|S_A|=2n|c_i|$, only linear with $n$. For instance, with $|k_{A,i}|=64,|c_i|=128$, we obtain a signature of length $|S_A|=2^{14}$, about two orders of magnitude shorter than in \cite{CidetalArxiv23}, using the same key length $L=nr=2^{12}$.

\begin{figure}[t]
\centering
\includegraphics[width=0.6\textwidth]{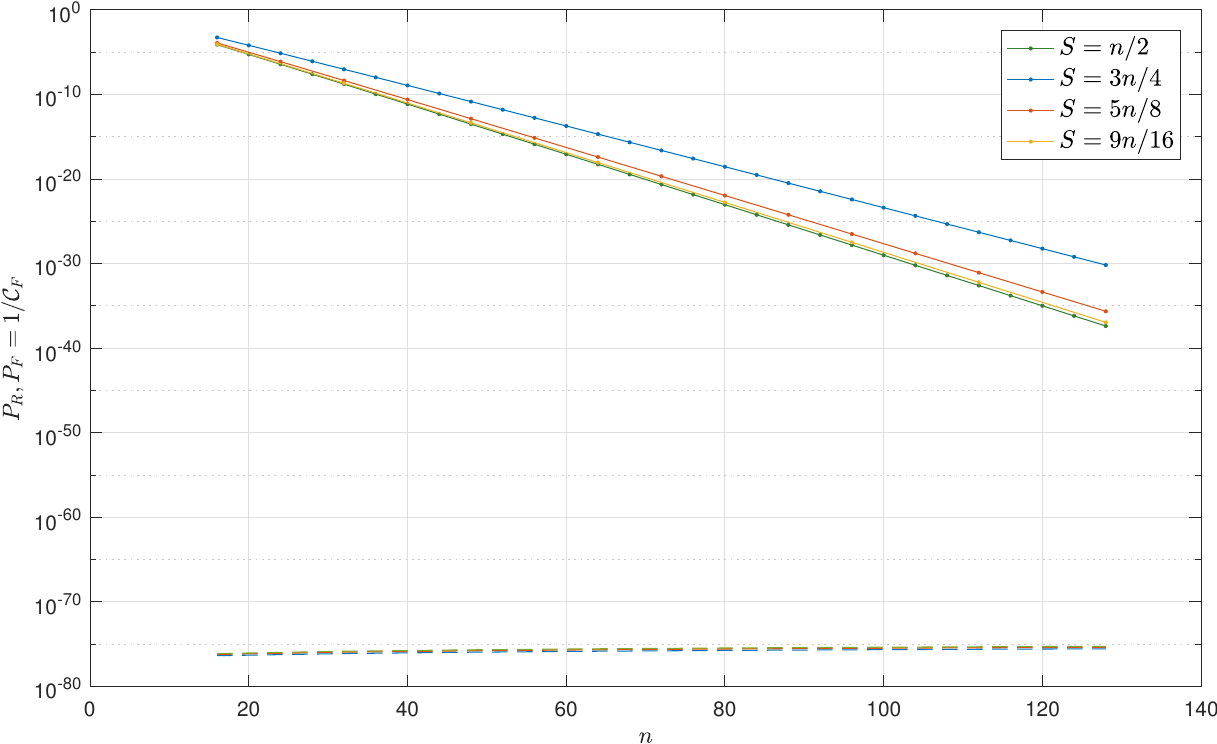}
\caption{Probability of repudiation and forgery as functions of $n$ for a GQaDS based on \emph{Poly1305} MAC, assuming $V_C=2S+1$, for various $S$.}
\label{fig:Prep Pfrg Poly1305 vs n}
\end{figure}

For instance, in Fig. \ref{fig:Prep Pfrg Poly1305 vs n} we plot $P_F$ and $P_R$ as functions of $n$ for GQaDS based on \emph{Poly1305},\footnote{Poly1305 is a UHF where the polynomial is defined on the finite field of the integers modulo the prime number $2^{130}-5$.} which is one of the most popular Carter-Wegman MAC proposed in \cite{Carter}. The curves are obtained by setting the threshold value $V_C=2S+1$, which minimizes $P_R$. The curves computed for various choices of $S$ confirm the optimal $S=n/2$. Note the \emph{Poly1305} key length $r=256$ guarantees $P_F<10^{-75}$, i.e. at least 225-bits security strength against forgery, whereas a minimum number of blocks $n= 84$ is mandatory to guarantee the minimum 80-bits security strength against repudiation  $\left(P_F<10^{-24}\right)$, in this \emph{probabilistic} setting. 

This final remark suggests the following variant, namely \emph{deterministic} GQaDS.

\subsection{Deterministic GQaDS}

As described in the previous section, GQaDS has both key length and signature length boosted by the large number of blocks $n \geq 64$. This number of blocks is necessary to lower the repudiation probability, as described by \eqref{eq - P[Rep|e]}, which is the probability that deliberate errors injected by a malicious Alice all occur in the signature portion that cannot be checked by Bob. In this case, Bob accepts the signature as valid, whereas Charlie will support Alice when she will attempt to repudiate her signature.

A simple way to overcome this risk  without the need of a large number of blocks, occurs in particular scenarios where the second verifier (Charlie) is a trusted third party and cannot be malicious. 
In this case, Bob's acceptance of Alice signature can be made contingent upon a positive check made by Charlie. This procedure is what we call \emph{deterministic GQaDS} since the inability for the signer to repudiate does not depend on a probabilistic argument, but rather on an \emph{assist} that the first verifier receives from the second. The advantage is that the number of blocks can be reduced to $n=2$, the minimum necessary to combat forgery attempts by Bob, since repudiation is impaired by the procedure itself. 

In fact, deterministic GQaDS  preserves the integrity, unforgeability and non-repudiation requirements of a digital signature and outputs a \textit{short} digital signature using QKD keys. 
Specifically, Alice divides $k_1$ and $k_2$ into just $n=2$ blocks of length $r$. This choice allows Alice to generate a concatenated key $k_A$ of $2 L= 4n = 1024$ bits, that she will separate in $2n=4$ blocks. Applying the Carter-Wegman MAC on each block of $k_A$ using the Poly1305 function, Alice's signature is composed by 512 bits only, much lower than that obtained in GQaDS.
In the GQaDS algorithm, after signing the message $m$, Alice sends the tuple $(m,S_A)$ to Bob, which performs the verification steps.
In deterministic GQaDS, Bob forwards the tuple $(m,S_A)$ to Charlie before starting the verification procedure locally. In this way, Bob’s and Charlie’s verification procedures will run at the same time and Bob must wait for Charlie's verification outcome to consider $m$ or $S_A$  as valid or not. In details, Bob accepts the signature if Charlie accepts it too. Otherwise, he rejects the message and the signature and considers them as invalid.

The drawback of this implementation is that, although Alice is still able to proceed autonomously in the signing phase, the verification requires the simultaneous presence of both verifiers. However, this can suitably fit many applications where one verifier (Charlie) plays an arbitrating role among many peers, i.e. many potential Alice or Bob, equipped with their reservoir of pre-shared QKD keys, quickly available from their local Key Manager in a QKD network.

\bigskip

    \section{Conclusions}
    In this paper, we propose GQaDS, a digital signature protocol whose IT security is based on QKD. The protocol overcomes the practical limitations of current QDS schemes, since it achieves the security strength recommended by NIST with practical key lengths. The scheme generalizes and optimizes a scheme previously presented in \cite{CidetalArxiv23}.  Our contribution includes a thorough analysis of the parameter selection, and the introduction of Carter-Wegman MACs to improve the scheme features. Moreover, in particular scenarios in which Charlie is an authority beyond suspicion, we propose deterministic GQaDS, a scheme which is based on a different approach and allows reducing further the key length.

\section*{Acknowledgment}

This work was co-funded by European Union - PON Ricerca e Innovazione 2014-2020 FESR /FSC - Project ARS01\_00734 QUANCOM.

The authors dedicate this paper to the memory of late Giovanni Schmidt, whose precious work on the subject has been of great inspiration for us. This paper would have been remarkably different without his leading contribution.

\end{document}